\documentclass[useAMS,usenatbib,letterpaper,10pt]{mn2e}
\usepackage{times}
\usepackage{epsfig}
\usepackage{natbib}
\usepackage{amssymb}
\usepackage[T1]{fontenc}
\usepackage{aecompl}
\usepackage[bookmarks,bookmarksnumbered,colorlinks=true, citecolor=blue, linkcolor=black]{hyperref}
\usepackage{color}
\title[The Formation of Massive, Compact Galaxies at $z=2$ in the Illustris Simulation]{The Formation of Massive, Compact Galaxies at $z=2$ in the Illustris Simulation}
\author[S. Wellons et al.]{Sarah Wellons$^{1}$\thanks{E-mail: swellons@cfa.harvard.edu}, Paul Torrey$^{2,3}$, Chung-Pei Ma$^{4}$, Vicente Rodriguez-Gomez$^{1}$, 
\newauthor Mark Vogelsberger$^{2}$, Mariska Kriek$^{4}$, Pieter van Dokkum$^{5}$, Erica Nelson$^{5}$,
\newauthor Shy Genel$^{1,6}\thanks{Hubble Fellow}$, Annalisa Pillepich$^{1}$, Volker Springel$^{7,8}$, Debora Sijacki$^{9}$, 
\newauthor Gregory Snyder$^{10}$, Dylan Nelson$^{1}$, Laura Sales$^{1}$, and Lars Hernquist$^{1}$\\
$^{1}$Harvard-Smithsonian Center for Astrophysics, 60 Garden St., Cambridge, MA 02138, USA\\
$^{2}$Massachusetts Institute of Technology, Cambridge, MA 02139, USA \\
$^{3}$TAPIR, Mailcode 350-17, California Institute of Technology, Pasadena, CA 91125, USA \\
$^{4}$University of California Berkeley, Berkeley, CA 94720, USA \\
$^{5}$Yale University, 260 Whitney Ave., New Haven, CT 06511, USA \\
$^{6}$Department of Astronomy, Columbia University, 550 West 120th Street, New York, NY, 10027, USA \\
$^{7}$Heidelberg Institute for Theoretical Studies, Schloss-Wolfsbrunnenweg 35, 69118 Heidelberg, Germany \\ 
$^{8}$Zentrum f\"ur Astronomie der Universit\"at Heidelberg, ARI, M\"onchhofstr. 12-14, 69120 Heidelberg, Germany \\
$^{9}$Kavli Institute for Cosmology, Cambridge, and Institute of Astronomy, Madingley Road, Cambridge, CB3 0HA, UK \\
$^{10}$Space Telescope Science Institute, 3700 San Martin Drive, Baltimore, MD 21218, USA}
\begin{document}

\maketitle

\label{firstpage}

\begin{abstract}
Massive, quiescent galaxies at high redshift have been found to be considerably more compact than galaxies of similar mass in the local universe.  How these compact galaxies formed has yet to be determined, though several progenitor populations have been proposed.  Here we investigate the formation processes and quantify the assembly histories of such galaxies in Illustris, a suite of hydrodynamical cosmological simulations encompassing a sufficiently large volume to include rare objects, while simultaneously resolving the internal structure of galaxies.  We select massive ($\sim10^{11}$ M$_\odot$) and compact (stellar half-mass radius $< 2$ kpc) galaxies from the simulation at $z=2$.  Within the Illustris suite, we find that these quantities are not perfectly converged, but are reasonably reliable for our purposes. The resulting population is composed primarily of quiescent galaxies, but we also find several star-forming compact galaxies.  The simulated compact galaxies are similar to observed galaxies in star formation activity and appearance.  We follow their evolution at high redshift in the simulation and find that there are multiple pathways to form these compact galaxies, dominated by two mechanisms: (i) intense, centrally concentrated starbursts generally triggered by gas-rich major mergers between $z \sim 2-4$, reducing the galaxies' half-mass radii by a factor of a few to below 2 kpc, and (ii) assembly at very early times when the universe was much denser; the galaxies formed compact and remained so until $z\sim2$.  
\end{abstract}


\section{Introduction}
\label{sec:intro}

Observations of the high-redshift universe have found a population of massive, quiescent galaxies \citep{VanDokkum2006, Kriek2006} which are substantially smaller and denser than their local counterparts at similar masses \citep{Daddi2005, Trujillo2006}.  Compact galaxies like these are rare in the local universe \citep{Trujillo2007, Toft2007, Buitrago2008, Cimatti2008, Franx2008, VanderWel2008, VanderWel2014, VanDokkum2008, Cassata2011, Damjanov2011, Szomoru2011, Bell2012}.  Their prevalence at high redshift and comparative scarcity at low redshift suggests that they either grow in size with time, cease to form at low redshifts, or (most likely) both.

\begin{table*}
 \centering
 \begin{minipage}{130mm}
  \caption{Parameters of the Illustris simulations at different resolution levels}
  \begin{tabular}{| l | c | c | c | c | c | c |}
  \hline
   Name  &  Volume & Number of DM & $\epsilon_{\rm baryon}$ & $\epsilon_{\rm DM}$ & $m_{\rm baryon}$ & $m_{\rm DM}$ \\
    & (Mpc$^3$) & particles + hydro cells & (pc, $z=2$) & (pc, $z=2$) & ($10^6$ M$_\odot$) & ($10^6$ M$_\odot$) \\
 \hline
 \hline
Illustris-1 & 106.5$^3$ & $2 \times 1820^3 \approx 1.2 \times 10^{10}$ & 473 & 473 & 1.26 & 6.26 \\
Illustris-2 & 106.5$^3$ & $2 \times 910^3 \approx 1.5 \times 10^{9}$ & 946 & 946 & 10.07 & 50.1 \\
Illustris-3 & 106.5$^3$ & $2 \times 455^3 \approx 1.9 \times 10^{8}$ & 1893 & 1893 & 80.52 & 400.82 \\
\hline
\end{tabular} \\
Note: The Plummer-equivalent gravitational softening length $\epsilon$ is constant in comoving units for $z > 1$ (but evolves in physical units).  Here we quote the most relevant value for this work, the physical softening length at $z=2$.  (After $z=1$, $\epsilon_{\rm baryon}$ is fixed to a physical size of 710 pc while $\epsilon_{\rm DM}$ continues to evolve.)
\label{tab:sims}
\end{minipage}
\end{table*}

It is possible that the observed growth in size of quiescent galaxies with time is due not to the growth of individual galaxies, but to the continual addition of larger galaxies to the quiescent population (progenitor bias, \citet{VanDokkum2001}, see also \citet{Carollo2013}). Progenitor bias alone may be an insufficient effect \citep{Belli2014, Belli2014a}, however, and there is evidence that high-redshift compact galaxies have grown into the centers of today's giant elliptical galaxies \citep{VanDokkum2010, VanDokkum2014}.  Dry (gas-poor) minor mergers which deposit material onto the outskirts of galaxies and puff up the existing system are one way to achieve this physical size growth \citep{Naab2007, Naab2009, Hopkins2009c, Hopkins2010, Bezanson2009, Oser2010, Oser2012, Johansson2012, Hilz2013, Feldmann2014}.  However, it is unclear whether the minor merger rate is high enough to produce the observed trend even with progenitor bias taken into account \citep{Newman2011}.  Galaxies may also expand adiabatically as a response to mass loss \citep{Fan2008, Fan2010}, but this process would only take place during early, active phases in their evolution.

The physical mechanism(s) by which compact galaxies form is also unclear, though the change in abundance from high to low redshift suggests that their formation is tied to the physical properties of the early universe.  Many of the proposed theoretical formation processes rely on the high mass fraction of cold gas in galaxies at that time, which can provide a source of dissipation that drives material inwards.  For example, major, gas-rich mergers where tidal torques drive gas rapidly to the centers of galaxies can provoke intense starbursts there, leaving a compact merger remnant behind \citep{Mihos1996, Hopkins2008, Wuyts2010}.  Disks rich in cold gas may also be susceptible to instabilities characterized by rapid, dissipative accretion and the formation of star-forming clumps which quickly sink to the galaxy center \citep{Dekel2009, Dekel2013}. 

These massive, compact galaxies are generally quiescent when observed, having already quenched their star formation activity at these high redshifts.  Potential star-forming compact progenitors of the $z\sim2$ quiescent compact galaxies have proven more elusive to observe.  Those which have been found have been heavily dust-obscured \citep{Gilli2014, Nelson2014} and may host active galactic nuclei (AGN) \citep{Barro2014}, making observations of them more difficult.

Observations can only offer a momentary glimpse into a galaxy's history.  Thus, we cannot know with certainty how any individual galaxy evolves, and observational studies are therefore limited to statistical comparisons of larger populations at different epochs.  Simulations of galaxy formation can provide the missing link between these discrete observational snapshots and directly connect galaxies at different redshifts.  Cosmological simulations with large volumes additionally possess a statistical advantage which allows galaxy populations such as compact ellipticals to arise ``naturally" without the specific contrivance of initial conditions which might favor their formation.

Herein, we describe how massive, compact galaxies form in Illustris, a set of cosmological simulations which trace the formation of structure in the universe from $z=127$ to the present day.  In Section \ref{sec:illustris} we describe the simulation methods employed in Illustris.  In Section \ref{sec:z2}, we describe the population of massive, compact galaxies present in the simulation at $z=2$ and make comparisons with properties of observed compact galaxies in the real universe. In Section \ref{sec:form}, we trace these compact galaxies back to higher redshift in the simulation and describe the methods by which they form, either by a central starburst or an early formation time.  In Section \ref{sec:converge} we examine the convergence of galaxy sizes and masses in Illustris and discuss the possible effects that resolution may have on the results.  Finally, in Section \ref{sec:discuss}, we summarize our findings, discuss their implications, and conclude.

\section{Illustris}
\label{sec:illustris}

\begin{figure*}
  \centering
  \includegraphics[width=2.1\columnwidth]{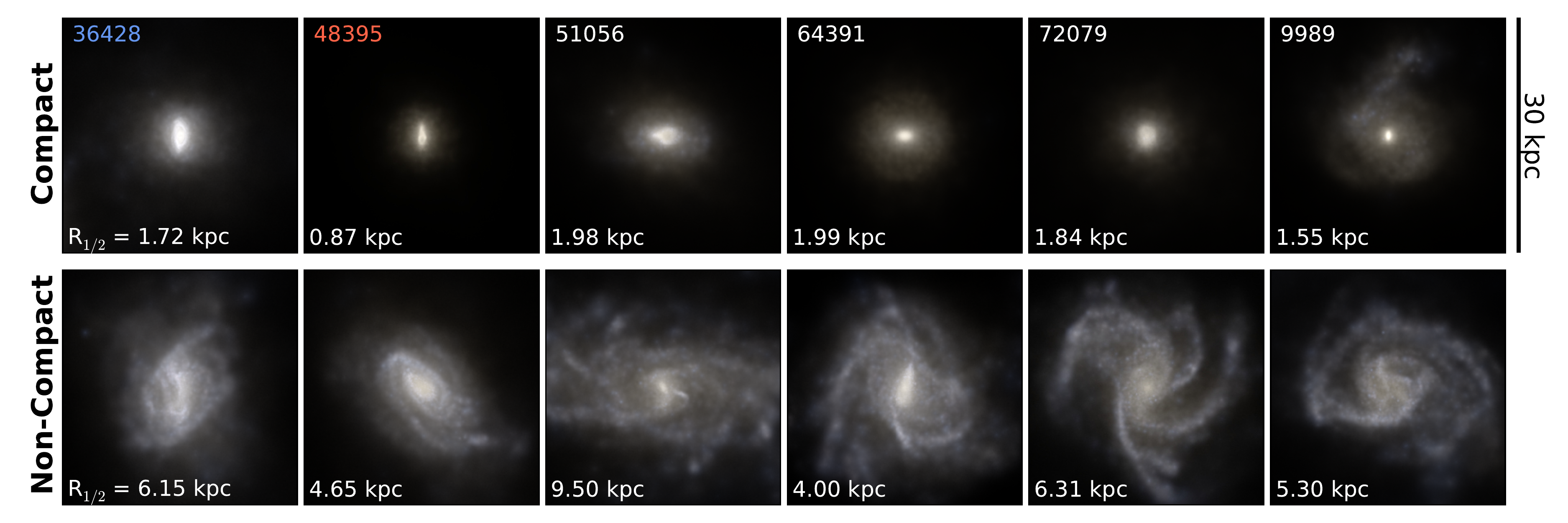}
  \caption{Images of compact (top row) and non-compact (bottom row) galaxies in Illustris-1, sampled from the population of 172 galaxies with stellar mass $1-3 \times 10^{11}$ M$_\odot$ at $z=2$.  The compact galaxies concentrate that mass into a dense central core, whereas the non-compact galaxies often have an extended star-forming disk.  Stellar half-mass radii $R_{1/2}$ are listed in the lower left corner of each panel.  Each image is a composite of stellar light in the SDSS g, r, and i bands.  The first two galaxies in the upper left, 36428 and 48395, are the same galaxies whose formation histories are examined in detail in Sections \ref{ssec:starburst} and \ref{ssec:early} respectively.}
  \label{fig:niceimages}
\end{figure*}

In Illustris, galaxies and other structures form in a box of comoving size (106.5 Mpc)$^3$ from cosmological initial conditions \citep{Vogelsberger2014, Vogelsberger2014a, Genel2014}.  In addition to gravitational forces, the simulation includes a hydrodynamic treatment of gas using the moving-mesh code {\sc AREPO} \citep{Springel2010}, which has demonstrated advantages on cosmological galaxy formation problems \citep{Vogelsberger2012, Keres2012, Torrey2012, Nelson2013} over traditional smoothed particle hydrodynamics codes.  The simulation also includes phenomenological models for processes important to galaxy formation which regulate the growth of stellar mass.  

Gas cools radiatively at a rate which is a function of gas density, temperature, and metallicity as well as the strength of radiation from AGN and the UV background.  When the gas reaches a hydrogen number density threshold of 0.13 cm$^{-2}$, it follows an effective equation of state for a two-phase gas as described by \citet{Springel2003}.  Stars form from this dense gas stochasically according to the Kennicutt-Schmidt relation \citep{Kennicutt1989} and return mass, metals, and energy to the gas via supernovae and stellar winds as they evolve.  Supermassive black holes (SMBHs) are seeded at early times in massive halos and grow with an accretion rate proportional to the density of the surrounding gas \citep{Springel2005}.  At high accretion rates, BH feedback follows a quasar-mode model which returns energy thermally to nearby gas.  At low BH accretion rates, radio-mode feedback driven by AGN jets drives gas out of galaxies mechanically with a duty cycle dependent on the BH mass and accretion rate \citep{Sijacki2007, Sijacki2014}.  The inclusion of feedback processes has been shown to be crucial to the formation of realistic disk galaxies (see e.g. \citet{Marinacci2014}).  See \citet{Vogelsberger2013} for a detailed description of the models, which have been tuned to match the galaxy stellar mass function and cosmic star formation rate density at $z=0$.

Using this treatment, the formation of galaxies from baryonic matter occurs simultaneously with that of the dark matter halos in which they are embedded, resulting in galaxy populations which evolve in a manner consistent with observations \citep{Torrey2014}.  The large volume of Illustris supplies a statistical sample of tens of thousands of galaxies while the resolution, though not approaching the level of detail accessible with ``zoom-in" simulations, is nevertheless capable of discerning galaxies' internal structure.  At $z=0$ (as well as $z=2$), our simulated galaxy population contains a variety of shapes and sizes: spiral and elliptical, compact and non-compact.  Thus, the balance that Illustris strikes between volume and resolution allows us to meaningfully ask what is different about the way that compact elliptical galaxies form, within the context of a diverse galaxy population.  

The simulation was run three times at different resolutions.  The parameters for each simulation can be found in Table \ref{tab:sims}. In particular, for the highest-resolution simulation (Illustris-1) from which our results are drawn, each baryonic resolution element has a mass of about $10^6$ M$_\odot$.  In the same simulation, the baryonic Plummer-equivalent gravitational softening length $\epsilon_{\rm baryon}$ is a comoving 1.42 kpc until $z=1$, at which point it is fixed at a physical size of 710 pc.  (At $z=2$, $\epsilon_{\rm baryon}$ has a physical size of 473 pc.)  The gravitational softening length for the gas is spatially adaptive, scaling with the cube root of the cell volume.  The convergence of sizes and masses between Illustris-1, -2, and -3 is discussed in Section \ref{sec:converge}.

\begin{figure}
  \centering
  \includegraphics[width=0.9\columnwidth]{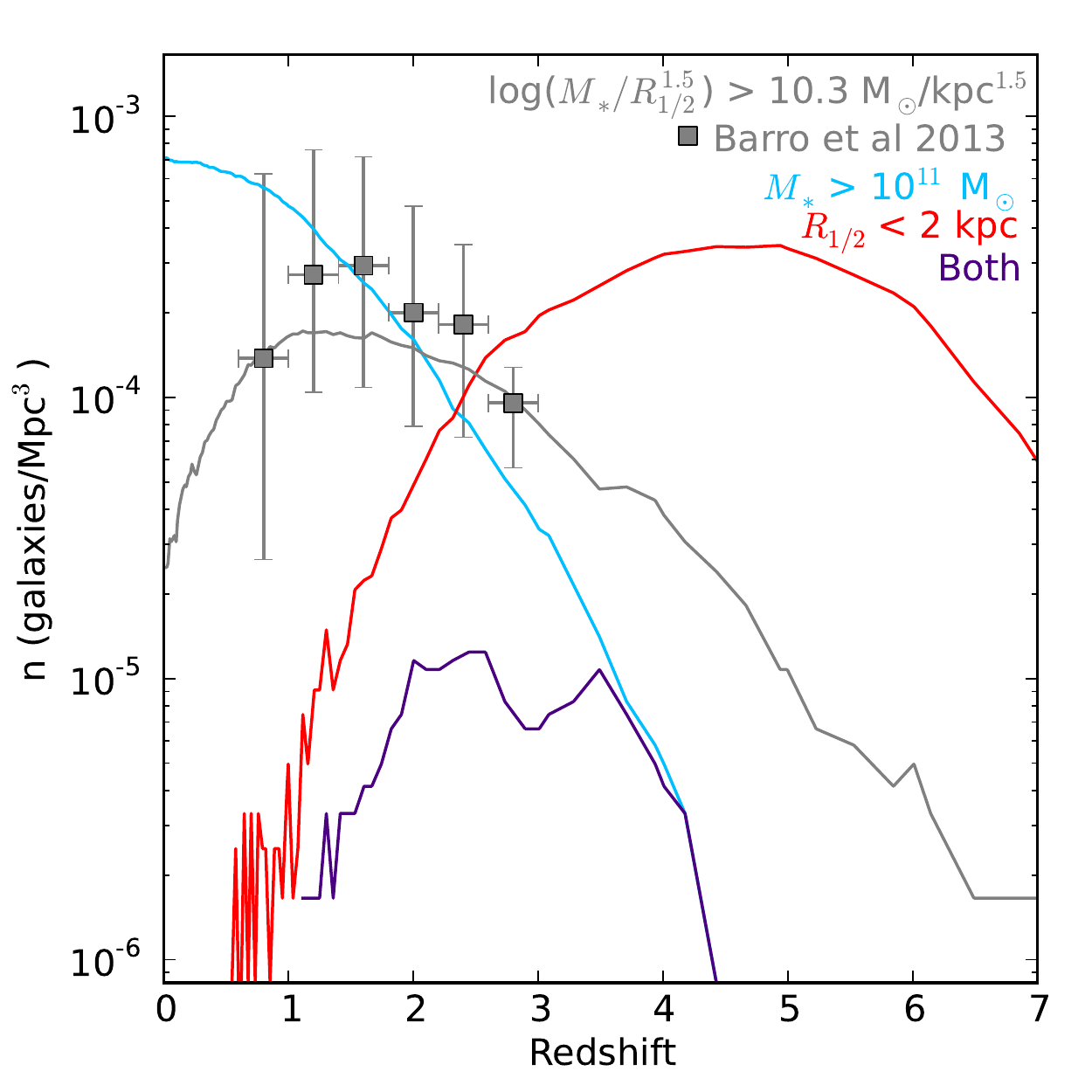}
  \caption{Comoving number densities of various galaxy populations.  Grey points indicate observational number densities of compact galaxies from \citet{Barro2013}, selected by applying the criterion log($(M/$M$_\odot$)/$(R$/kpc$)^{1.5}) >$ 10.3 where $M$ and $R$ are stellar mass and half-light radius respectively.  The number density for the same population in Illustris is shown by a grey line (where we use a projected stellar half-mass radius as an estimate of the half-light radius), and shows reasonable agreement between the simulation and observations.  The two criteria for the selection of compact galaxies in the remainder of the paper are shown individually in red (stellar half-mass radius $<$ 2 kpc) and blue (stellar mass $> 10^{11}$ M$_\odot$), and in combination in purple.  At $z=2$, 14 galaxies meet these criteria.}
  \label{fig:numberdensities}
\end{figure}

Gravitationally-bound substructures of baryons and dark matter (`subhalos') are identified using the {\sc SUBFIND} algorithm \citep{Springel2001, Dolag2009}.  The baryonic components of each subhalo - gas, stars, and black holes - comprise its associated galaxy.  The two properties of each galaxy with which we are primarily concerned in this paper are its stellar mass $M_*$, which is the sum of the masses of its star particles, and its stellar half-mass radius $R_{1/2}$, which is the radius at which a sphere centered on the galaxy's most-bound particle encloses half of its stellar mass.

Subhalos are connected between simulation snapshot outputs using {\sc SUBLINK} (Rodriguez-Gomez et al., submitted), which uses the common ownership of individual particles to determine a subhalo's line of descent.  A merger takes place when two or more subhalos have the same descendant.  The main progenitor of each branch of the merger tree is chosen to be the one with the ``most massive history" \citep{DeLucia2007}.  We find the higher-redshift progenitors of our compact galaxies and identify merger events in their pasts by traversing this merger tree.

\begin{figure*}
  \centering
  \includegraphics[width=1.8\columnwidth]{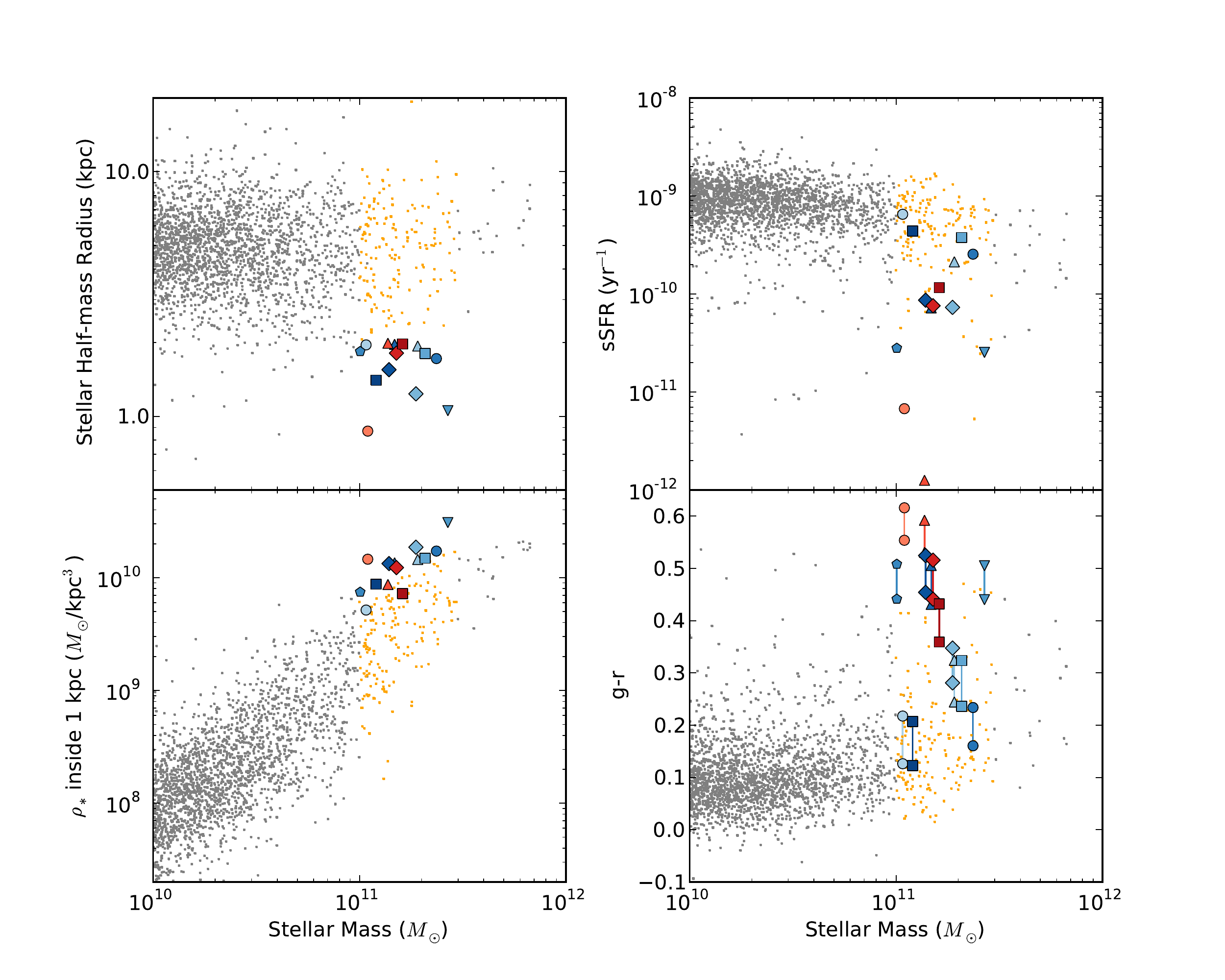}
  \caption{Properties of compact galaxies (colored symbols) at $z = 2$, in the context of the overall galaxy population (grey and orange points) in the simulation.  The colors of the symbols for the compact galaxies represent the different formation mechanisms discussed in Section \ref{sec:form}. (Orange points represent non-compact galaxies in the same $z=2$ mass bin, as in Figure \ref{fig:mrscatter}.)  The criteria used to select the compact galaxies -- stellar mass greater than $10^{11}$ M$_\odot$ and stellar half-mass radius under 2 kpc -- can be seen in the top left panel.  These large stellar masses and small sizes produce high stellar densities in the inner kpc of the galaxies (bottom left panel).  Many have quenched and fallen off the star formation main sequence (top right panel), and consequently appear redder than average (bottom right, colors measured in rest frame).  4-5 compact galaxies, however, are still on the star formation main sequence and are blue.  For each of the compact galaxies, we also show the color when a simple model for dust reddening is included (\citet{Charlot2000}, upper half of each pair).}
  \label{fig:z2}
\end{figure*}

The simulation was run to the present day at $z=0$, but in this paper we examine the state of simulation at $z=2$ since compact ellipticals are a high-redshift phenomenon.  Images of a selection of galaxies present in the simulation at that time are shown in Figure~\ref{fig:niceimages}.  All of the galaxies shown have stellar masses of $1-3 \times 10^{11}$ M$_\odot$.  In this mass bin (containing 172 galaxies in all) we find a range of morphologies.  In addition to the extended, actively star-forming disks (bottom row), we find some galaxies which are elliptical and compact, similar to those observed in the real universe.  Some examples of these compact galaxies are shown in the top row of Figure \ref{fig:niceimages} and are the focus of our study.

\section{Properties of Simulated Compact Galaxies at $z=2$}
\label{sec:z2}

\subsection{Number densities and selection criteria}
\label{ssec:selection}

\begin{figure*}
  \centering
  \includegraphics[width=2.1\columnwidth]{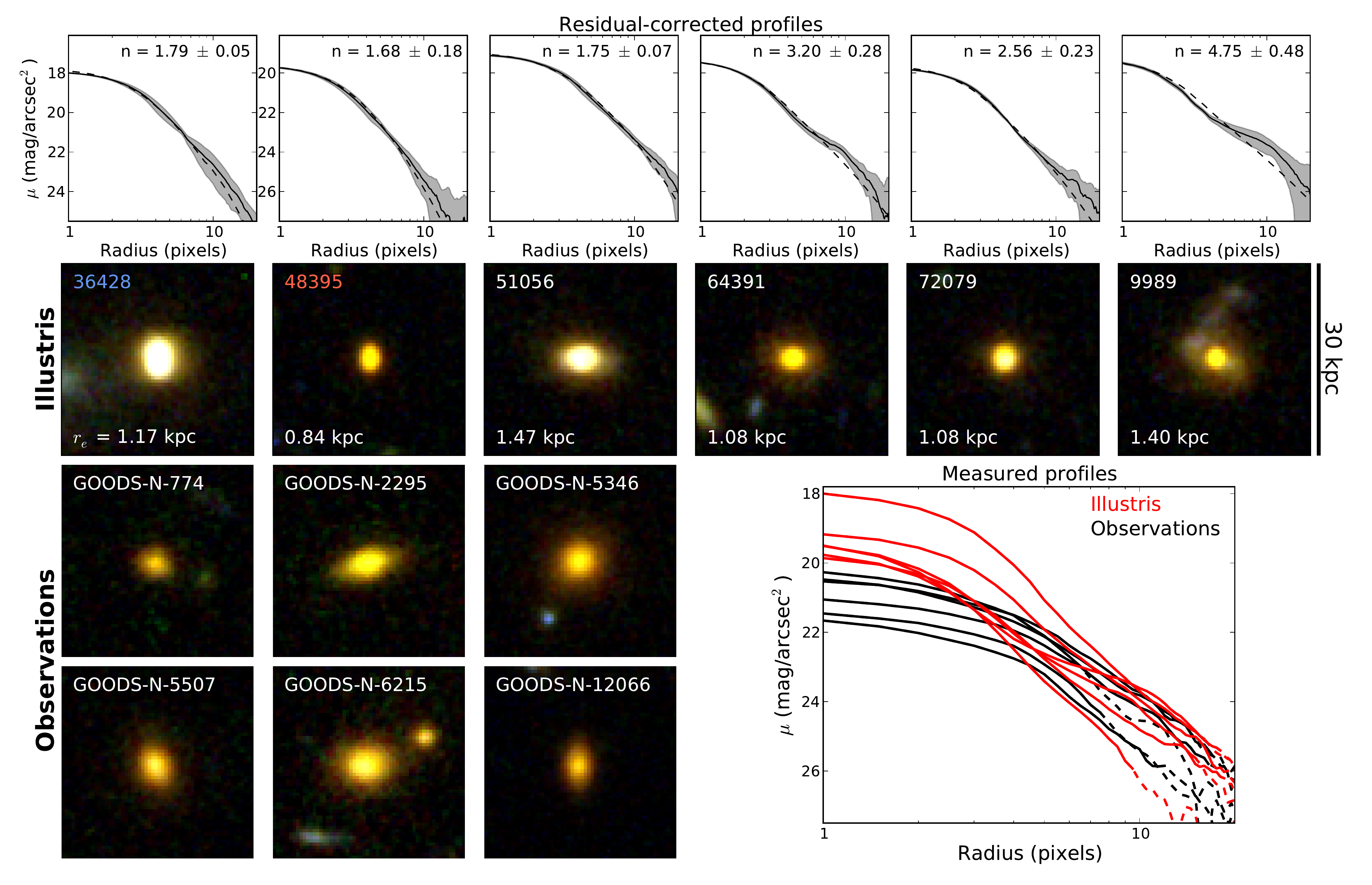}
  \caption{Comparison of mock HST images of simulated compact galaxies (second row) with images of real compact galaxies from the 3D-HST catalogs in the CANDELS fields (bottom left, \citet{Skelton2014}).  The mock images from Illustris were developed using techniques by \citet{Torrey2014a}, described in Section \ref{ssec:observations}.  Both sets of images are color composites of bands F125W, F160W, and F606W from WFC3, with identical stretch and limits applied.  The bottom right panel shows surface brightness profiles measured directly from each of the 12 panels in band F160W.  (Dashed lines indicate that the standard deviation of the pixels at that radius exceeds their median value.)  The profiles of the Illustris galaxies have similar surface brightnesses to the observations at large radii but are brighter in the center (possibly because the mock images do not include dust).  In the top row, a dashed line shows profiles measured from images of the best single-Sersic fit to each of the mock observations, obtained using {\sc GALFIT} \citep{Peng2010}.  The Sersic index $n$ and effective radius $r_e$ for these fits are listed in the first and second row respectively.  A residual correction is then applied in the solid lines by adding back on the profile measured from the residual image that remains after the model is subtracted out, as in \citet{Szomoru2012}.  The shaded regions around the solid lines indicate the standard deviation of these residuals at a given radius.  Despite the multiple-component nature of the underlying stellar surface density profiles, there are only two cases (64391 and 9989) where the surface brightness profiles are clearly distinguishable from a single-component model.  In these cases the second components do not actually belong directly to the compact cores, but were accreted well after the core's formation.}
  \label{fig:mockimages}
\end{figure*}

Observationally, compact galaxies have a number density $2 \times 10^{-4}$ Mpc$^{-3}$ at $z=2$, where ``compact" is as defined by \citet{Barro2013} to be galaxies which satisfy log($(M/$M$_\odot$)/$(R$/kpc$)^{1.5}) >$ 10.3, where $M$ is stellar mass and $R$ is half-light radius in the HST/WFC3 $H$-band.  When we select for compact galaxies at $z = 2$ in Illustris using the same criterion\footnote{As an approximation for the projected half-light radius, we use the 3D stellar half-mass radius times a projection factor of 0.75.  This projection factor was chosen by measuring and comparing the 2D and 3D stellar half-mass radii of the compact systems from the simulation, and finding 0.75 to be the median ratio of the two.} we find a similar number density of $1.5 \times 10^{-4}$ Mpc$^{-3}$.    The evolution in number density of these systems, shown in Figure \ref{fig:numberdensities}, is also similar to observations, rising at high redshift to peak around $z=1-2$ and falling thereafter. 

For a detailed study of the formation histories of the most massive and compact galaxies in the simulation, we narrow the ``compact" selection criteria further by requiring stellar masses above 10$^{11}$ M$_\odot$ and stellar half-mass radii below 2 kpc.  The number densities of galaxies meeting these individual mass and size criteria are shown in blue and red respectively.  These conspire in such a way that the number density of compact galaxies meeting both criteria (shown in purple) reaches its peak around $z=2-3$.  At higher redshift, galaxies which have already assembled that much stellar mass are rare, and at lower redshift, they have larger physical sizes.

\subsection{Star formation rate and color}
\label{ssec:sfr}

At $z=2$, we find a sample of 14 massive and compact galaxies which meet the mass and size criteria described in Section \ref{ssec:selection}.  These are shown in the context of the overall galaxy population in Figure \ref{fig:z2} with the mass-radius selection criteria depicted in the top left panel.  In addition to their sizes, there are several other properties which distinguish the compact galaxies from the general population.  Most of them (about 2/3) have quenched and fallen off the star formation main sequence, possessing specific star formation rates below $2 \times 10^{-10}$ yr$^{-1}$.  By selecting for small galaxies, we have preferentially also selected for quiescent galaxies.  In the mass range of $1-3 \times 10^{11}$ M$_\odot$, quiescent galaxies have a median half-mass radius of 2.7 kpc while star-forming galaxies have a median half-mass radius of 5.2 kpc.  These compact quiescent galaxies comprise the majority of observational data, but we also find in the simulation 4 or 5 examples of the observationally-elusive compact star-forming galaxies which have not yet fully quenched.

The lower star formation rates of the compact galaxies imply that most of their stars were formed at higher redshift, and these older stellar populations cause the compact galaxies to also appear redder than is typical.  The lower right panel of Figure \ref{fig:z2} shows the g-r colors for all the $z=2$ galaxies, calculated by adding together the light from the galaxies' constituent star particles (each of which contains a population of stars formed at the same time).  Each compact galaxy has an additional point (the higher, redder one) which includes a simple dust model \citep{Charlot2000} which assumes a baseline amount of reddening for every star particle, plus additional reddening for very young stars still in their birth clouds.  This particular model does not take into account the high column densities along lines of sight through compact galaxies' cores, so we expect that with a more realistic model even the most actively star-forming compact galaxies would be heavily obscured and appear significantly redder.

\subsection{Comparison with Observations}
\label{ssec:observations}

In the second row of Figure \ref{fig:mockimages}, the same $z=2$ compact galaxies from Figure \ref{fig:niceimages} are shown as they would be ``observed" with the Hubble Space Telescope using the mock observation methods developed by \citet{Torrey2014a}. In short, the SEDs of the star particles which comprise each galaxy were redshifted to $z=2$ and integrated over the F125W, F160W, and F606W bands from HST's Wide Field Camera 3 (WFC3) in pixels of 0.06''\footnote{At $z=2$, 0.06'' is about 0.5 kpc, similar to the gravitational softening length $\epsilon$ in the simulation.} using the {\sc SUNRISE} code \citep{Jonsson2006}. The resulting image was then convolved with a Gaussian PSF with FWHM 0.16'', and coadded with an observational noise background.  This procedure is similar to techniques that have been used to compare zoom-in simulations with HST observations (e.g., \citet{Moody2014, Snyder2014}).  For comparison, images of real observations of compact galaxies from the 3D-HST catalogs in the CANDELS fields \citep{Skelton2014}, in the same bands and with the same stretch, are shown to the bottom left.  

The simulated and observed galaxies have similar appearances in terms of their sizes and colors.  The mock observations do not include dust, however, and so are brighter and less red in the central regions where we expect high dust column densities.  This difference in central brightness is especially visible in the bottom right panel showing the surface brightness profiles as measured directly from the images in the F160W band.  The profiles from the Illustris mock images are shown in red, while those from real observations appear in black.  Dashed lines indicate that the noise is comparable to the magnitude of the line at that radius.  In the outer regions where the obscuring effect of dust is not as strong, the observed and simulated profiles cover a similar range of surface brightness.

The gravitational softening used in the simulation may affect the profiles of these galaxies on the scale of $\epsilon = 0.5$ kpc, so we do not put too much weight on their exact shapes and values.  (See Section \ref{sec:converge} for a discussion of the effects of softening.)  It is worth noting, however, that in general the underlying stellar mass surface density profiles have a two-component shape similar to those of simulated compact merger remnants by \citet{Hopkins2008} and \citet{Wuyts2010} and local quiescent galaxies \citep{Hopkins2009b, Hopkins2009d}, with a dense inner component with low Sersic index $n_s~\lesssim~1$ and a diffuse outer component with higher $n_s$.  This stands in contrast to observations by \citet{Szomoru2012} and \citet{Williams2014}, who measure only a single component for high-redshift compact galaxies and do not detect the presence of any diffuse wings.  We perform a similar analysis on our mock images to see whether their wings in surface density manifest as detectable wings in surface brightness.

For each of the simulated galaxies shown in Figure \ref{fig:mockimages}, the best-fit model to a single-Sersic profile was found using {\sc GALFIT} \citep{Peng2010}.  These fits are shown with a dashed line in each panel of the top row.  The Sersic indices $n_s$ and effective radii $r_e$ for the first five (good) fits lie in the ranges of $n_s = 1.7 - 3.5$ and $r_e = 0.8 - 1.5$ kpc.  The presence of any other components in the galaxies' structure is then accounted for using residual-correction, where the residuals from the fit are added back onto the best-fit model.  This technique has been shown by \citet{Szomoru2010} to capture the true profile robustly, even against poor fitting in the first step.  The black line in each panel indicates this residual-corrected profile, and the shaded area the standard deviation of the residuals. 

The galaxy furthest to the right (9989) demonstrates the ability of residual-correction to capture deviations from a single-component fit.  9989 is in the middle of an interaction at $z=2$ and has recently accreted a gas-rich satellite, giving it tidal structure and a clear second component.  To a lesser extent, this is also true of 64391, which possesses a small, misaligned stellar disk acquired well after the formation of the compact core.  Any deviations from a single-Sersic fit for the remaining four galaxies lie within the standard deviation of the noise.  In all six cases, no secondary components directly belonging to the compact cores are clearly detected, despite their presence in the underlying surface density profiles.  Thus, the existence of wings in simulated compact merger remnants should not necessarily preclude their consideration as potential progenitors of compact galaxies.

\section{Formation Histories}
\label{sec:form}

\begin{figure}
  \centering
  \includegraphics[width=0.95\columnwidth]{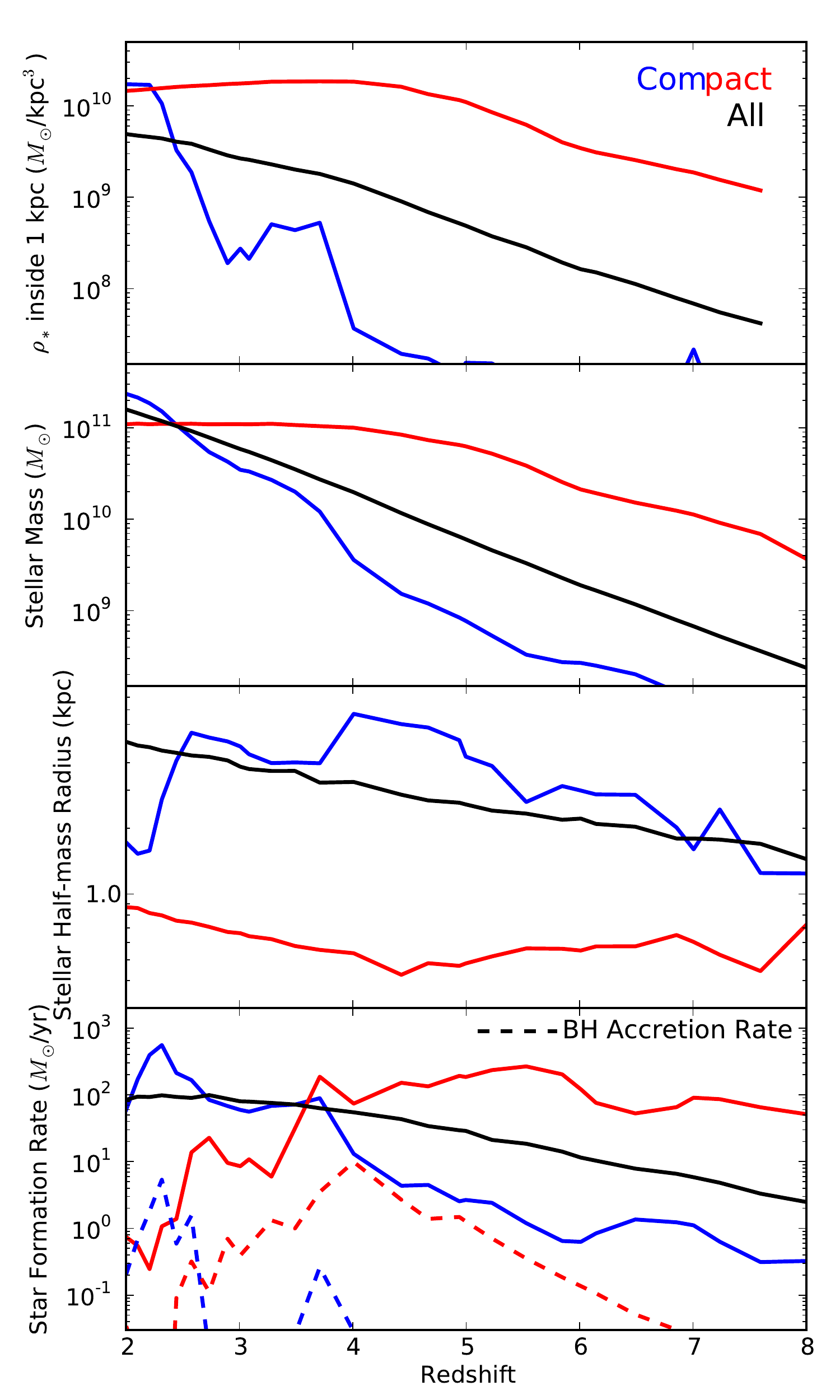}
  \caption{Evolution in redshift of stellar mass density in the central kpc (first panel), stellar mass (second panel), stellar half-mass radius (third panel), and star formation rate (last panel).  The final panel also contains a dashed line following the accretion rate of the galaxies' central black holes, which tends to track star formation rate in the compact systems.  Blue and red lines depict the two compact galaxies described in Sections \ref{ssec:starburst} and \ref{ssec:early} respectively.  The former owes its compactness primarily due to an intense central starburst brought on by a gas-rich major merger at $z \sim 2.5$, while the latter formed its stellar mass at very early times when the scale factor was small.  The black line in each panel shows for comparison the average evolution of 172 galaxies with $1-3 \times 10^{11}$ M$_\odot$ at $z=2$ (including both compact and non-compact galaxies).}
  \label{fig:twotracks}
\end{figure}

Galaxies build up their stellar mass by a combination of in-situ star formation and accretion from, or mergers with, other galaxies.  We trace our compact galaxies through the simulation's merger trees back to their progenitors at higher redshifts to see how they assembled.  We find two primary mechanisms which produce the compactness observed at $z = 2$: central starbursts, and early formation.  The evolution of stellar mass, half-mass radius, star formation rate, and central density for a prototypical compact galaxy for each of these two formation channels are shown in Figure \ref{fig:twotracks} in blue and red, respectively.  The black line depicts the average evolution of all galaxies (compact and non-compact) which have stellar masses of $1-3 \times 10^{11}$ M$_\odot$ at $z=2$.  

The following two subsections describe in detail the formation of the two galaxies in Figure \ref{fig:twotracks}.  These galaxies have been selected to illustrate the two compact formation mechanisms that we find in the simulation.  In Section \ref{ssec:wholesample}, we then extend the discussion to the entire compact sample.

\subsection{Example 1: central starburst}
\label{ssec:starburst}

The first galaxy (shown in blue) is under-massive compared to other galaxies with similar masses at $z=2$ for most of its lifetime.  It grows rapidly during an episode of violent, intense star formation exceeding a rate of 500 M$_\odot$/yr around $z = 2.5$.  The majority of the galaxy's stellar mass is formed during this episode, during which the stellar half-mass radius drops from 6 kpc to under 2 kpc.  This rapid reduction in half-mass radius occurs because the burst of star formation is concentrated at the galaxy's core, prompted by a gas-rich major merger.  Tidal torques from the merger drive much of the cold gas to the center, and the high gas densities there induce heavy star formation, resulting in a compact merger remnant \citep{Mihos1996, Hopkins2008, Wuyts2010}.  A similar but less intense merger and associated starburst event is visible earlier in the galaxy's history, around $z=4$.

During each event, the accretion rate of the central supermassive black hole peaks sharply.  The same high central gas densities which produce strong star formation also feed black hole growth \citep{DiMatteo2005, Springel2005}, so we expect more AGN activity during this period \citep{Hopkins2006}.  The presence of an AGN could make such star-forming compact galaxies more difficult to observe, especially in combination with dust obscuration.  Even at the time of peak star formation, this galaxy would be heavily dust-obscured and so may appear as a sub-millimeter galaxy, another class of galaxies that have been suggested as possible progenitors for the compact population \citep{Toft2014}.

\subsection{Example 2: early formation}
\label{ssec:early}

In contrast to the dramatic starburst events which characterize the evolution of the first galaxy, the second (shown in red) has a relatively quiet formation. It is among the most massive galaxies in the simulation at high redshift, forming stars at a rate that far exceeds its peers.  Nearly all of its $10^{11}$ M$_\odot$ in stellar mass is in place by $z = 4$.  Around that time, the star formation rate decreases as a result of gas depletion and stellar feedback and finally quenches completely from thermal quasar-mode feedback driven by high accretion rate onto the central SMBH.  After this galaxy falls off the star formation main sequence, radio-mode AGN feedback inhibits further accretion and maintains low star formation rates.

Half of this galaxy's stellar mass was assembled by $z = 4.8$.  In comparison, an average galaxy in the simulation with a similar stellar mass at $z = 2$ will have assembled half of that mass by $z = 3.2$.  During the earlier epoch when the compact galaxy formed, the universe itself was much more dense  ($a = 0.17$ vs. $a = 0.24$), and so the galaxy naturally formed with a smaller size \citep{Mo1998, Shull2014}.  Between $z = 4$ and $z = 2$, after its star formation terminates, the galaxy is quiescent and does not undergo any significant merger events, thereby retaining the small size with which it formed.  

\begin{figure*}
  \centering
  \includegraphics[width=2\columnwidth]{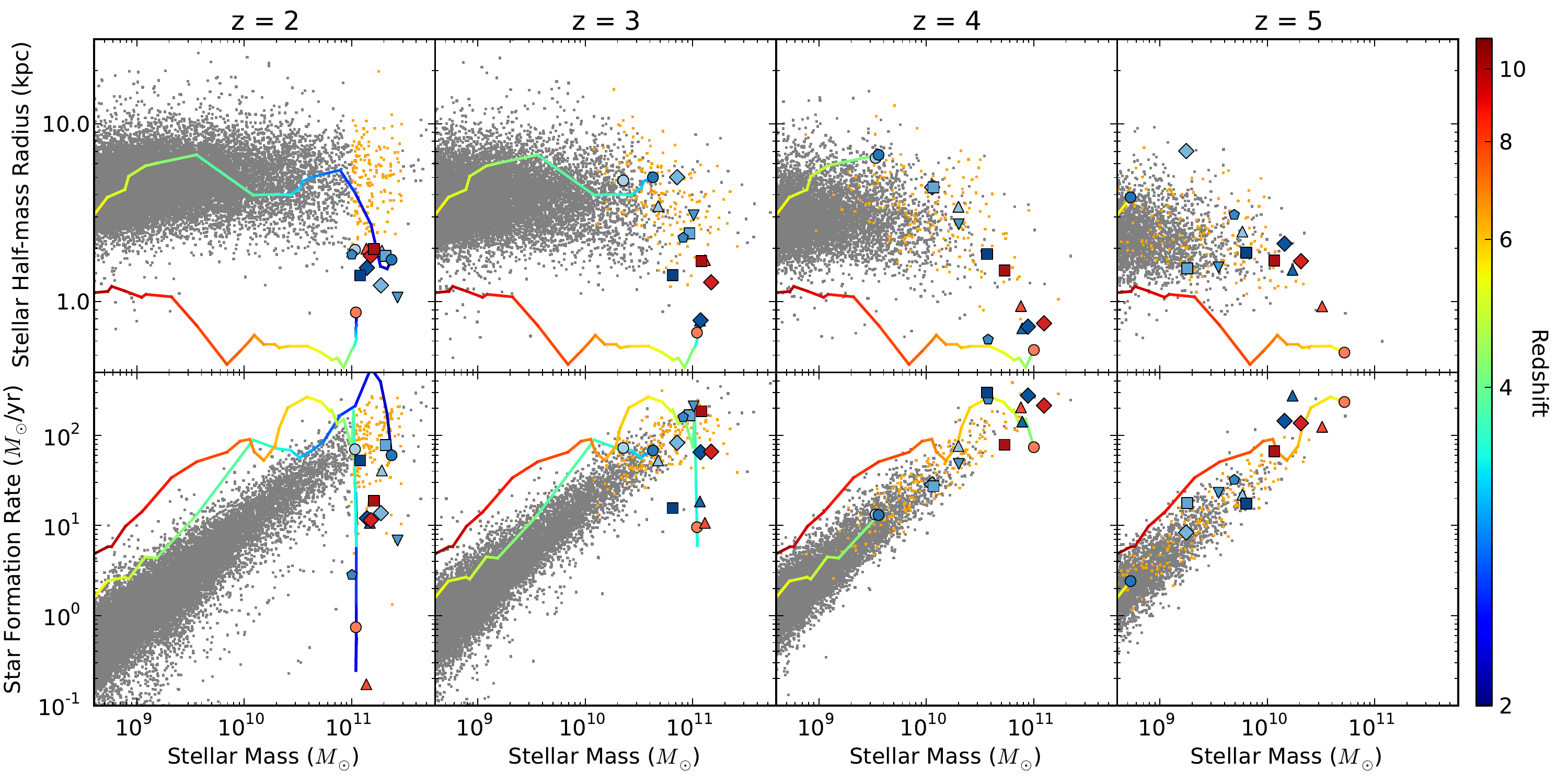}
  \caption{Stellar mass against stellar half-mass radius (top) and star formation rate (bottom) of the entire galaxy population in the simulation at four different redshifts, with the compact galaxies and their progenitors picked out in colored symbols.  Orange points indicate non-compact galaxies which reach the same range of stellar mass at $z = 2$.  Two lines in each panel follow the formation of the two galaxies from Figure \ref{fig:twotracks}, with the color of the line at a given point indicating the redshift when the galaxy reached that location.  Each compact galaxy (or its progenitor) is represented by a symbol whose color represents its dominant formation mechanism.  The variety in formation channels is evident here -- some galaxies form their stars at high redshift and subsequently fall off the star formation main sequence (red symbols), and others lag behind, catching up quickly during a burst of star formation (blue symbols).  (The different shades within red and blue are not significant other than to differentiate the galaxies.)}
  \label{fig:mrscatter}
\end{figure*}

\begin{figure*}
  \centering
  \includegraphics[width=2\columnwidth]{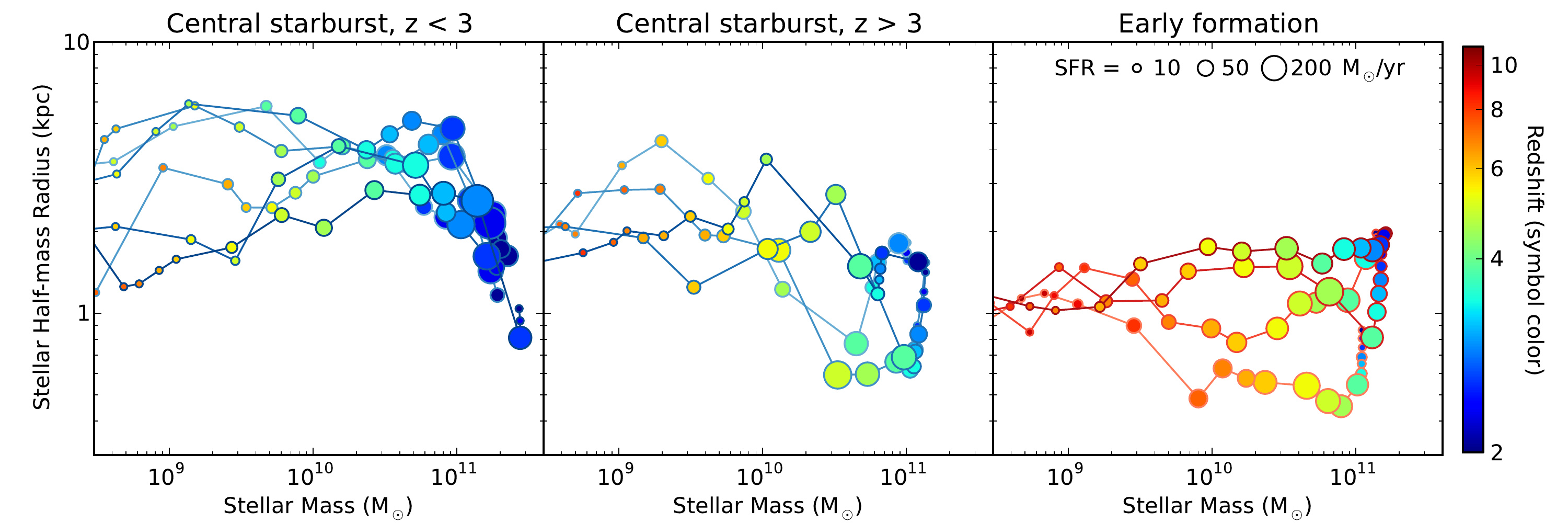}
  \caption{Tracks across the M-R diagram for each of the 14 galaxies, separated into those which formed early (right panel), and those which experienced a starburst before (middle panel) or after (left panel) $z=3$.  Symbol color indicates the redshift when the galaxy reached that point, and the symbol size indicates its star formation rate at that time.  In the left and middle panels, all the galaxies experience a drop in radius coincident with a high SFR, indicating a compact starburst.  In the right panel, each galaxy had high SFR at high redshift, reaching its $\sim 10^{11}$ M$_\odot$ stellar mass very early.}
  \label{fig:mrtracks}
\end{figure*}

\subsection{Entire compact sample}
\label{ssec:wholesample}

The two galaxies described in the previous subsections take very different tracks across the mass-radius diagram.  These tracks can be seen in the top-left panel of Figure \ref{fig:mrscatter}, where the color of the line at a given point indicates the redshift when the galaxy reached that location.  As in Figure \ref{fig:twotracks}, it is evident that the first galaxy experiences a quick drop in radius shortly before $z=2$, while the second assembles its mass at high redshift and has a small size throughout.  

Each of the 14 compact galaxies we identified at $z = 2$ forms in a way that resembles one of these two archetypal paths, or some combination thereof.  The remaining panels in Figure \ref{fig:mrscatter} show snapshots of the masses, radii, and star formation rates of their progenitors at $z$ = 2, 3, 4, and 5.  Many of the compact galaxies (those shown in blue) form in a manner similar to the one described in Section \ref{ssec:starburst}, lagging behind at high redshift and forming most of their stars in a burst of concentrated star formation brought on by a merger or interaction before $z = 2$.  These events bring their half-mass radii below the 2 kpc threshold in a short, sudden drop.  Others (red shapes) owe their compactness primarily to early formation, similar to the galaxy from Section \ref{ssec:early}: they are some of the most massive galaxies at high redshift, having assembled most of their mass early.  They quench early as well, being among the first to fall off the star formation main sequence.  As a result of their high-$z$ assembly, they form with small physical sizes and maintain a small half-mass radius until $z = 2$, undisturbed by major mergers.  

Figure \ref{fig:mrtracks} shows the individual mass-radius tracks for all 14 galaxies, separated into three panels by formation mechanism and time.  The rightmost panel shows the tracks for the galaxies colored in a shade of red in Figures \ref{fig:z2} and \ref{fig:mrscatter}, those which are compact primarily because they formed early.  The sizes of the symbols correspond to star formation rate; in this panel the four reached their peak star formation rate around $z=4-5$ and have had half-mass radii $R_{1/2} < 2$ kpc throughout.  Note that some galaxies are affected by both mechanisms.  One galaxy in the rightmost panel formed early, but also had a merger event around $z=4$ which decreased its size.  Thus, though the two mechanisms we have described here are distinct, the galaxies themselves form a continuum as they are affected more or less by each one.

The middle and left panels of Figure \ref{fig:mrtracks} show the mass-radius tracks for the galaxies represented by blue points in Figures \ref{fig:z2} and \ref{fig:mrscatter}, which experienced a central starburst.  For clarity of display, they have been separated according to whether the primary starburst occurred before or after $z=3$.  (As in the example from Section \ref{ssec:starburst}, some galaxies experience multiple events.)  The objects in the middle panel are analogous to a galaxy at $z=3.35$ recently found by \citet{Marsan2014} to have a stellar mass of $3 \times 10^{11}$ M$_\odot$, an effective radius under 1 kpc, a low star formation rate, and indications of a starburst prior to $z=4$, as well as an AGN.  All the galaxies in these two panels exhibit a rapid drop in half-mass radius coincident with a peak in star formation rate.  The majority of these starburst events are precipitated by a major merger (mass ratios of 3:1 or lower), though a few of the triggering mergers are more minor with mass ratios as high as 5:1.  An exception is one of  the galaxies in the left panel of Figure \ref{fig:mrtracks}, which does not have a single triggering merger.  Rather, the disk of cold gas around the galaxy appears to be driven by heavy accretion or multiple small mergers to buckle and collapse of its own accord.

The same high densities of gas that prompt intense central star formation also feed the supermassive black holes at the galaxies' centers, producing AGN.  In the model employed for SMBHs in Illustris \citep{Springel2005}, the black hole accretion follows an Eddington-limited Bondi-Hoyle-Lyttleton rate which goes as $\dot{M}_{\rm BH} \propto M_{\rm BH}^2 \rho/c_s^3$ for a stationary black hole where $\rho$ and $c_s$ are the density and sound speed of the surrounding gas.  As a result, the black holes at the centers of the compact galaxies are more massive than those residing in non-compact galaxies, by about a factor of two.  Because the large black hole mass and the compactness of the galaxy are both consequences of the high central gas density, the two go hand-in-hand.  Relatively large masses of the black holes are ubiquitous among the compact ellipticals in the simulation, regardless of the details of their formation.  These periods of high black hole accretion rate coincide with the periods of high star formation rate, similar to recent Herschel observations of distant radio galaxies by \citet{Drouart2014}.  Therefore, it is possible that star-forming compact galaxies are preferentially contaminated by AGN, making them more difficult to observe and accounting for their observational scarcity. 

\begin{figure}
  \centering
  \includegraphics[width=\columnwidth]{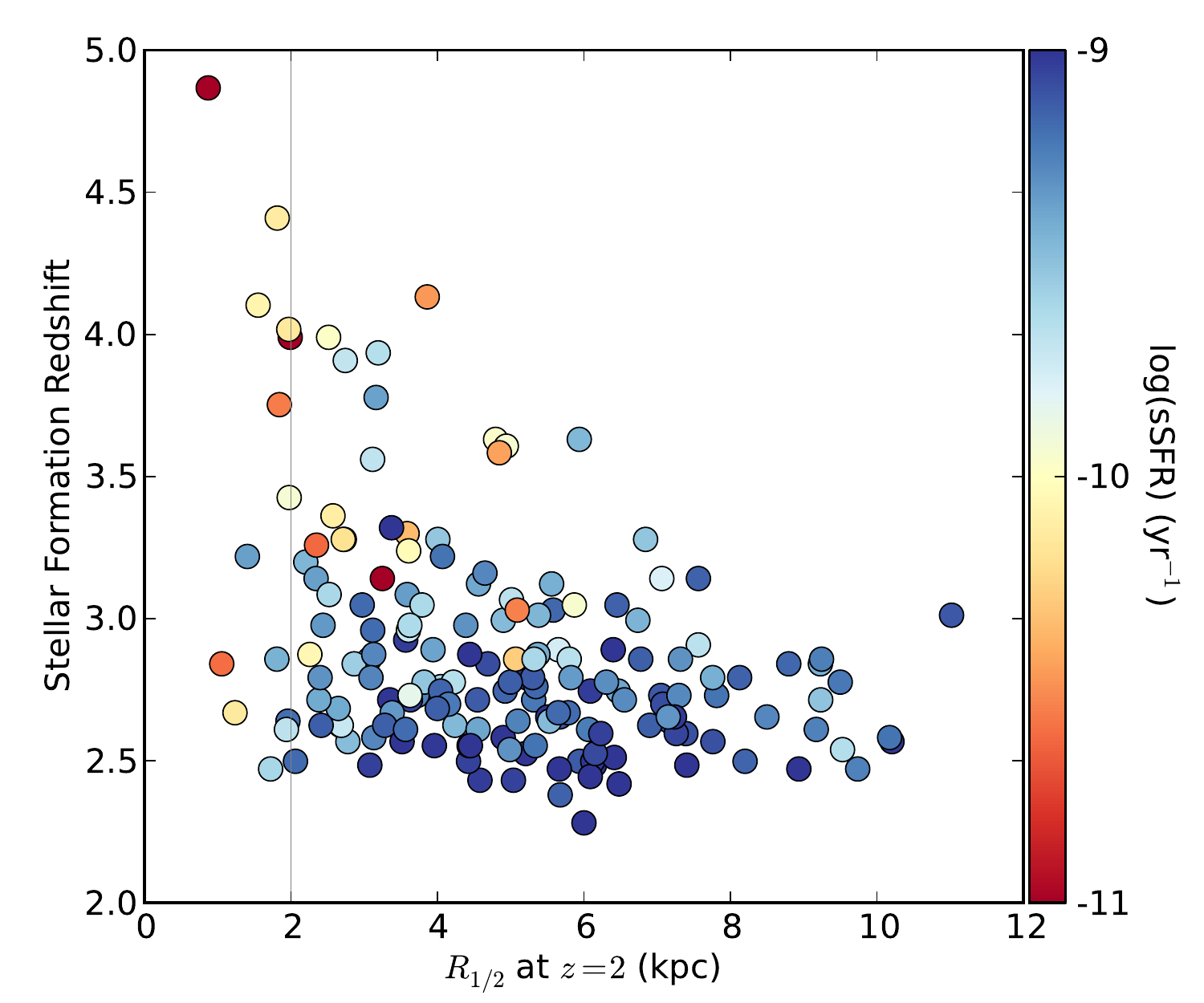}
  \caption{Stellar formation redshift for all galaxies of stellar mass $1-3 \times 10^{11}$ M$_\odot$ at $z=2$ in Illustris-1, against their $z=2$ half-mass radii.  Here the stellar formation redshift is defined as the median formation redshift of all the star particles the galaxy possesses at $z=2$, weighted by their initial mass at the time of formation.  Galaxies which form at higher redshift are preferentially small, while later-forming galaxies may be small or extended.  Each marker is also colored according to the galaxy's specific star formation rate at $z=2$.  Those which are quenched are preferentially smaller and older, explaining the high quiescent fraction among the compact galaxies with $R_{1/2} <$ 2 kpc (grey line).}
  \label{fig:formtimesize}
\end{figure}

Figure \ref{fig:formtimesize} shows how the sizes of galaxies throughout the $1-3 \times 10^{11}$ M$_\odot$ sample depend on the redshift at which they formed. Here, we define the formation redshift to be the median formation redshift of all the star particles in the galaxy (for simplicity, both in-situ and ex-situ), weighted by the particles' initial masses.  The smallest galaxies with sizes $\lesssim$ 2 kpc can form at any redshift prior to $z=2$, as we have seen in the last two examples.  Similarly, galaxies which formed recently may form at any size within 1 kpc $\lesssim R_{1/2} \lesssim$ 10 kpc.  However, the maximum size that a galaxy can have is a function of redshift, as galaxies which form half of their stellar mass at earlier times are constrained to be smaller. Therefore, we do find in our compact sample several galaxies whose compactness is a symptom of their early formation time. 

\begin{figure*}
  \centering
  \includegraphics[width=1.8\columnwidth]{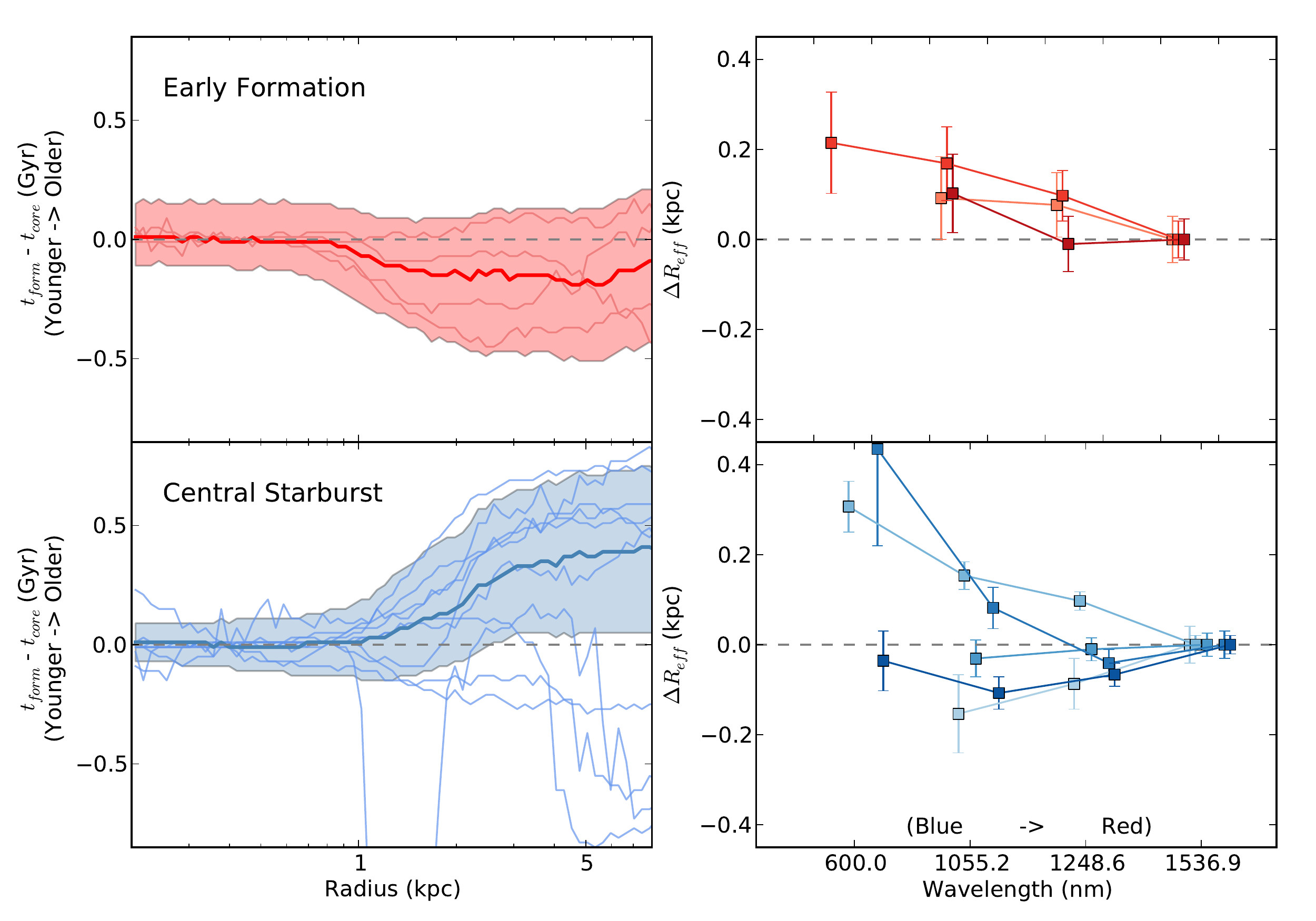}
  \caption{ {\it Left panels}:  Stellar age as a function of radius in the Illustris compact galaxies, in comparison to the age of the core.  Thin lines show the median stellar age at a given radius for individual galaxies, with the median age of the star particles inside 0.7 kpc subtracted out.  Thick lines indicate the median stellar age difference, and shaded regions the 30-70 percentile range, for the stack of all the galaxies in each panel.  
  {\it Right panels}: Change in effective radius with wavelength, measured from Sersic fits to mock observations of the Illustris compact galaxies in HST WFC3 bands F606W, F105W, F125W, and F160W (in order on the x-axis from short to long wavelength).  The effective radius of each galaxy in band F160W has been subtracted out.  Measurements are only shown for cases where a reasonable fit could be found, excluding e.g. galaxies undergoing interactions or bands with low signal-to-noise.  
Those galaxies which formed very early (as in Section \ref{ssec:early}) appear in the top panels and those which experienced a central starburst (as in Section \ref{ssec:starburst}) appear in the bottom panels.  Early-forming galaxies tend to have an old core and slightly younger outskirts due to a small amount of ongoing star formation, and thus have a red to blue color gradient which can be seen as larger sizes at shorter wavelengths.  Cores which formed in starbursts tend conversely to be the youngest component, embedded in an older component which is a remnant of the merger progenitors.  These then have blue to red color gradients and smaller sizes at shorter wavelengths.  A few exceptions to these trends do exist.  Some galaxies have accreted satellites since the core's formation or are undergoing an interaction at $z=2$ and so possess an additional, much younger outer component.  The two galaxies in the lower-right panel with $\Delta R_{\rm eff} > 0$ also clearly fall outside the expected trend in color gradient.  These are very recent starbursts that still have high SFR in both components at $z=2$ and so appear large at short wavelengths, but for them the color gradient is of secondary importance since their formation mechanism is immediately identifiable by their SFR.  In general, gradients in color reflect gradients in stellar age, and so could be an observational indicator of a compact galaxy's formation mechanism.}
  \label{fig:colorgradient}
\end{figure*}

By further distinguishing galaxies according to their sSFR at $z=2$, we find that the upper left corner of the stellar formation time - size plane is mostly occupied by quiescent galaxies.  Here we find again that the majority of our compact galaxies at $z=2$ are quiescent, even though star-forming objects can be found among the same compact sample.  Galaxy sizes in Illustris will be further explored in Genel et al. (in preparation).

\subsection{Observational differences}
\label{ssec:obsdiff}

How might one distinguish between these formation mechanisms observationally?  The very recent starbursts which have not yet completely quenched are easily picked out by star formation rate or color.  Those compact galaxies which underwent their starburst events at higher redshifts $z > 3$, however, have been quiescent long enough by $z=2$ to possess the same low star formation rates and redder colors as those which formed early.  Thus these simple criteria are insufficient to separate the formation mechanisms.

Nor do we find a significant difference between the two mechanisms in the shape of the galaxies' stellar mass distributions.  As mentioned in Section \ref{ssec:observations}, in general the compact galaxies possess two-component stellar density profiles regardless of origin.  In the case of the starbursts, a compact component forms on top of the preexisting structure of the merger progenitor(s).  Many of the early-forming galaxies also pick up a second component from the small amount of ongoing star formation at larger radii that can occur even after the compact core itself has quenched.  (Galaxy 64391 in Figure \ref{fig:mockimages} is one example of this phenomenon.)

\begin{figure*}
  \centering
  \includegraphics[width=2\columnwidth]{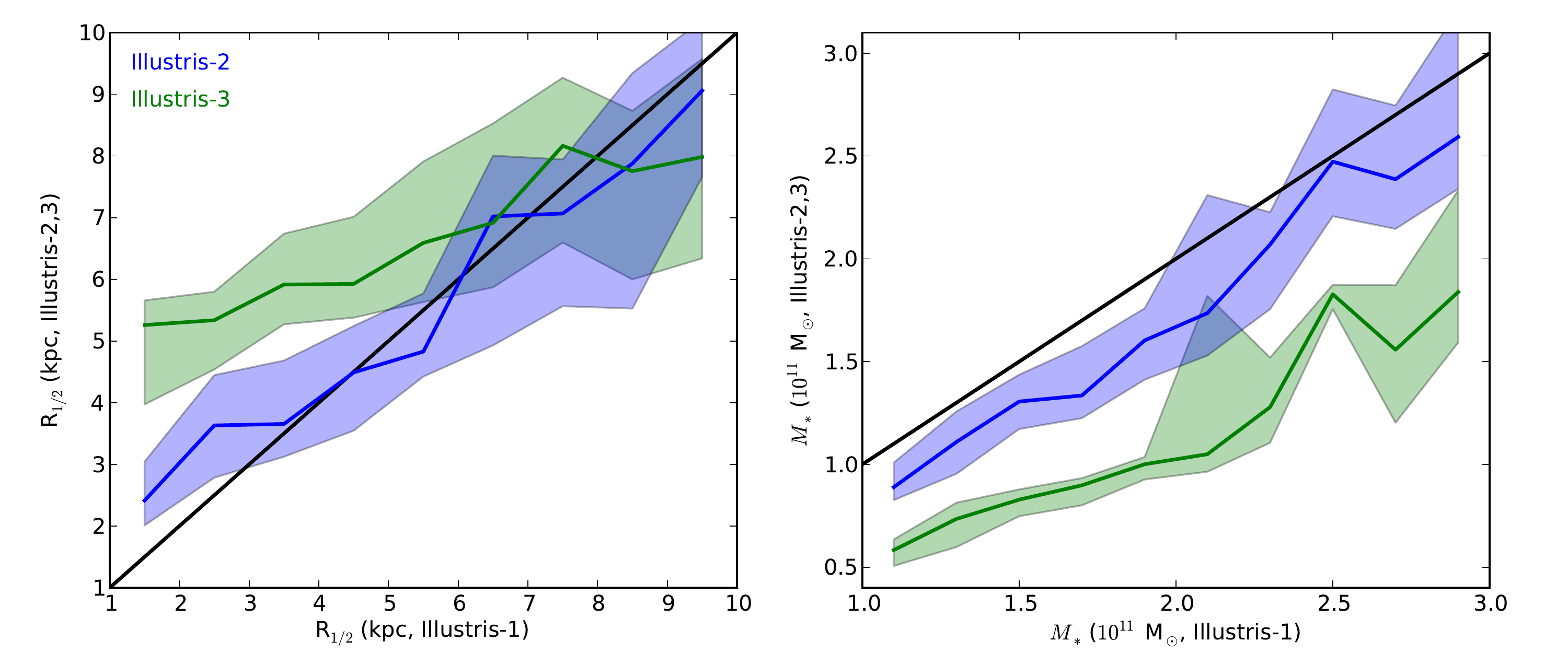}
  \caption{Convergence of sizes and masses of Illustris galaxies.  All the galaxies with stellar masses of $1-3 \times 10^{11}~M_\odot$ at $z=2$ in Illustris-1 (172 in total) are binned according to their stellar half-mass radii (left panel) or stellar mass (right panel).  Each of these galaxies has an analog in the lower-resolution simulations.  The solid lines show the Illustris-1 size (black) plus the median deviation from that size for the Illustris-2 (blue) or -3 (green) analogs in that bin.  Shaded areas represent the 25th-75th percentiles of the distributions.  Both mass and size are converging, masses increasing with resolution and sizes decreasing with resolution. The galaxy sizes are converged down to about 3 kpc between Illustris-1 and -2.  Thus, all the galaxies we identify as ``compact" in Illustris-1 would remain compact at higher resolution (they would be smaller and/or more massive, if anything), but our sample is most likely incomplete.}
  \label{fig:analogsizes}
\end{figure*}

Though the stellar density profiles themselves are similar, the differing order in which the two components form can manifest as gradients in stellar age (and therefore color) which run in differing directions.  For the starburst-produced galaxies, the central component is the youngest (and bluest) part, while for the early formers the central component is the oldest (and reddest) part.  The thin lines in the left panels of Figure \ref{fig:colorgradient} show the stellar age profiles of each of the compact galaxies, in relation to the age of their compact core.  Those which formed early (top panel) tend to have outer components which are younger than the core, and those which experienced a central starburst (bottom panel) tend to have outer components which are older than the core.  A few of the galaxies experienced minor mergers following the formation of the compact core, and so possess an additional outer component of very young material accreted in these mergers.  When the galaxies in each panel are stacked together, the difference in stellar age gradient between the two formation mechanisms is very clear.  The solid line and shaded area in each panel show the median and 30-70th percentile range of the distribution of the difference between stellar and core formation time in these stacks.

These gradients in stellar age should appear observationally as gradients in color.  We test this idea by fitting single-Sersic profiles to mock observations (see Section \ref{ssec:observations}) in the WFC3 bands F105W, F125W, F160W, and F606W .  If a galaxy has a color gradient, its measured effective radius will change as a function of wavelength.  (E.g., galaxies which are redder in the center will appear smaller at longer wavelengths).  

The results of this analysis are displayed in the right-hand panels of Figure \ref{fig:colorgradient}, which shows the best-fit effective radii in each band relative to the F160W value.  As expected, those galaxies which formed early tend to have larger sizes at shorter wavelengths, indicating a red to blue color gradient.  The results for the galaxies that experienced a central starburst are somewhat more ambiguous.  Several display the expected blue to red color gradient, but there are a couple of notable exceptions.  Both of these galaxies' central starbursts occurred quite recently around $z=2.5$ (in fact, one is the example from Section \ref{ssec:starburst}), and both still remain on the star formation main sequence at $z=2$, not yet having fully quenched.  In these cases, the remnants of the progenitor disks are still quite blue and so they are also detectable at shorter wavelengths.  The high star formation rates in these galaxies at $z=2$ mark them as unambiguously as starbursts, however, so their apparent color gradients are unlikely to cause confusion.  

Thus, we suggest that measuring compact galaxies' color gradients may provide observers with an idea of how they were formed.  We note, however, that in real observations the transformation between stellar age gradients and color gradients could be complicated by the presence of dust and/or AGN, the effects of which were not included in this analysis.  

\section{Convergence and effects of resolution}
\label{sec:converge}

Because compact galaxies are by definition some of the smallest in the simulation, they are potentially subject to resolution effects.  The gravitational softening employed in the simulation smooths out systems on the scale of the Plummer-equivalent softening length $\epsilon_{\rm baryon}$ (hereafter simply $\epsilon$ because we concern ourselves primarily with the baryons).  From $z=2-3$, the gravitational softening has a physical scale $\epsilon = 473 - 355$ pc, which means that our upper bound of 2 kpc on the sizes of compact systems is only $(4-6) \epsilon$ during that time and the sizes may not be well converged.  To explore the effects of resolution on both the sizes and masses of Illustris galaxies, we compare the Illustris-1 galaxies with their Illustris-2 and -3 counterparts.  

Each $\sim 10^{11}$ M$_\odot$ galaxy in Illustris-1 has an analog in the lower-resolution Illustris-2 and -3 simulations, where $\epsilon$ is twice and four times as large respectively.  In Figure \ref{fig:analogsizes}, the masses and sizes of these analogs are shown in comparison with Illustris-1.  In each panel, all the Illustris-1 galaxies with stellar masses of $1-3 \times 10^{11}$ M$_\odot$ at $z=2$ are binned according to their size or mass in the high-resolution simulation.  Each bin (containing about 20 galaxies each) then has a distribution of lower-resolution analog sizes or masses whose median value and 25th-75th percentile regions are shown in the Figure.  

From this, it is apparent that as resolution improves and these properties approach their converged values, galaxies' stellar masses increase while their stellar half-mass radii decrease.  Between Illustris-1 and -2, we find that the sizes of galaxies are converged down to about 3 kpc.  In our ``compact" range of 2 kpc and below, however, the sizes are not yet converged and could be smaller with better resolution.  Similarly, the stellar masses of Illustris-1 galaxies may not be converged, and could be possibly be larger.

Both of these potential changes with higher resolution (smaller sizes, larger masses) act in the direction of increasing galaxies' compactness.  We are therefore confident that the galaxies we have selected to be compact in Illustris-1 would be at least as compact in a higher-resolution simulation, so the selection for compactness is robust.  We note, however, that our resolution tests imply that the sample is likely incomplete.  The numerical uncertainties on our simulated galaxies' sizes and masses are mirrored by observational uncertainties on the same properties of compact galaxies in the real universe.  Our aim is not to try and match these parameters exactly, but rather to provide physical insight on how compact systems form in our simulation and may also form in the real universe.  

Though our main finding (that the most compact galaxies in our simulation owe their small sizes to early formation and/or compact starbursts) is robust against changes in resolution, some caution is warranted in interpreting certain other results.  For example, in Section \ref{ssec:selection}, we measured the number density of compact galaxies according to log($(M/$M$_\odot$)/$(R$/kpc$)^{1.5}) >$ 10.3 for comparison with observations (grey line in Figure \ref{fig:numberdensities}).  For galaxies with stellar masses below 10$^{10.9}$ M$_\odot$ or so, the (3D) size threshold for qualifying as ``compact" according to this definition lies below 3 kpc (where Illustris-1 and -2 sizes diverge).  Therefore, it is possible that this estimate of number density is missing some lower-mass galaxies which have not converged to their true sizes.  (Inflation of the number density by even a factor of 3 at $z=2$, however, would still leave it within the observational error bars.)  The galaxy profiles shown in Figure \ref{fig:mockimages} may be also affected by gravitational softening.  The exact shapes and values of these profiles should therefore not be relied upon.  The fact that the observed surface brightness profiles seem to possess only a single component while the underlying mass density profiles have a double-component structure, however, is significant regardless of whether the density profiles are physically believable in detail.

\section{Summary and Conclusion} 
\label{sec:discuss}

The combination of high mass resolution and large volume provided by Illustris allows, for the first time, the study of galaxy populations like compact ellipticals in a fully cosmological context.  Their small number density ($\sim 10^{-4}/$Mpc$^3$ at $z=2$) requires a large volume to achieve reasonable statistics, and their small size requires a mass resolution which can capture galaxies' internal structure.  The coincident evolution of baryonic and dark matter throughout Illustris' (106.5 Mpc)$^3$ volume produces tens of thousands of galaxies with a variety of masses and morphologies.  At $z=2$, the simulation produces both star-forming, disky galaxies and quiescent, spheroidal galaxies.  We searched the simulation at that redshift for massive, compact galaxies and found a population whose characteristics are broadly similar to observed compact galaxies.  They are dominated by quiescent galaxies (though we also find some star-forming compact galaxies), their number density peaks around $z=1-2$ and drops thereafter, and mock images of these galaxies have similar appearances to observational images.  

We then traced the compact galaxies in the simulation across cosmic time to determine how their formation differs from that of normal-sized galaxies.  We find two major mechanisms to be responsible for compactness at $z=2$: central starbursts and early formation.  In the first scenario, large inflows of gas trigger an intense burst of centralized star formation, usually driven by a gas-rich merger.  In the second scenario, galaxies which form their stars rapidly at high redshift assemble at a time when the universe was very dense, and so have small sizes throughout their lifetimes.  

Both of these mechanisms rely on conditions which are unique to the high-redshift universe - namely, its density and richness of cold gas.  Galaxies which assemble at lower redshifts form with larger sizes due to the expansion of the universe, so the first mechanism becomes unavailable with time.  The size of a merger remnant depends, among other things, on the gas content of the progenitor galaxies.  As the abundance of cold gas in galaxies decreases, so too does the likelihood of a gas-rich major merger to occur and produce a central starburst, making the second formation pathway less efficient with time as well.  With fewer and fewer new additions to the compact population, these galaxies become increasingly rare at lower redshift as its $z = 2$ members are disturbed, consumed by other galaxies, or simply grow larger by accretion from other galaxies and their surroundings.  The fate of compact galaxies at low redshift in Illustris will be discussed in detail in a subsequent paper.

Neither of these phenomena (high-redshift accretion and mergers) are unique to compact galaxies.  Other galaxies in the same mass range underwent similar processes but are nevertheless non-compact at $z=2$ (such as some which form early but then accrete satellites, or others which experience major mergers with gas fractions or orbital parameters that resulted in non-compact remnants).  Thus it seems that high-redshift compact elliptical galaxies do not represent a special class unto themselves, but are rather a tail of the smooth distribution that is a natural consequence of all the different combinations of physical processes galaxies may undergo.  

\section*{Acknowledgments}

SW is supported by the National Science Foundation Graduate Research Fellowship under grant number DGE1144152.  LH acknowledges support from NASA grant NNX12AC67G and NSF grant AST-1312095.  We thank Pascal Oesch for helpful discussions and guidance on generating the mock observations presented in this paper.

\label{lastpage}

\end{document}